
\input phyzzx

\baselineskip=16pt
\titlepage
\title{Equivalence between light-cone and conformal gauge in
the two-dimensional string}
\bigskip
\author{C. Arvanitis$^1$\footnote\dag
{c.arvanitis@ic.ac.uk; supported by the European Community under Human
Capital and Mobility Grant No. ERBCHBICT941235}
and R. Tzani$ ^{1, 2}$\footnote\ddag
{r.tzani@ic.ac.uk; supported by the European Community under Human
Capital and Mobility Grant No. ERBCHBICT930727}}
\address{ 1. Imperial College, Theoretical Physics Group,\break
Prince Consort Road, London SW7 2BZ, G.B.}
\address{ 2. The University of Ioannina, Division of Theoretical
Physics, 45110 Ioannina, GREECE} 
\bigskip
\abstract{Aiming towards understanding the question of the discrete
states in the light-cone gauge in the theory of two-dimensional
strings with a linear background charge term, we 
study the path-integral formulation of the theory. In particular, 
by gauge fixing Polyakov's path-integral expression for the 2-d
strings, we show that the 
light-cone gauge-fixed generating functional is the same as the conformal
gauge-fixed one and is critical for the same value of the background
charge (Q=2 $\sqrt 2 $). 
Since the equivalence is shown at the generating functional level, one
expects that the spectra of the two theories are the same. The zero
modes of the ratio of the determinants are briefly analyzed and it is 
shown that only the constant mode survives in this formulation. 
This is an indication that the discrete states may lie  
in these zero modes. This 
result is not particular to the light-cone gauge, but it holds for the
conformal gauge as well. 
}
\vfill
\eject

\def\NP{{\it Nucl. Phys.\ }}
\def\PL{{\it Phys. Lett.\ }}

\def\PRD{{\it Phys. Rev. D\ }}

\def\PRL{{\it Phys. Rev. Lett.\ }}
\def\CMP{{\it Comm. Math. Phys.\ }}

\def\Mod{{\it Mod. Phys. Lett.\ }}

\REF\gros{D. J. Gross, I.R. Klebanov and M.J. Newman, \NP {\bf B350}
(1991) 21.}
\REF\klebanov{I.R. Klebanov and A.M. Polyakov, \Mod {\bf A6} (1991)
3273.}
\REF\dani{U. H. Danielsson and D. H. Gross, \NP {\bf B366} (1991) 3.}
\REF\lian{B. H. Lian and G. J. Zuckerman, \PL {\bf B254} (1991) 417.}
\REF\witten{E. Witten, \NP {\bf B373} (1992) 187.}
\REF\bouw{P. Bouwknegt, J. McCarthy and K. Pilch, \CMP {\bf 145}
(1992) 541.}
\REF\wien{S. Weinberg, \PL {\bf B156} (1985) 309, {\it Strings and
Superstrings}, proceedings of the Jerusalem Winter School for
Theoretical Physics, Jerusalem 1985, edited by T. Piran and
S. Weinberg (World Scientific).} 
\REF\smith{E. Smith, \NP {\bf B382} (1992) 229; \PL {\bf B305} (1993) 344.}
\REF\rud{R.E. Rudd, {\it Ph.D. Thesis} Princeton University (1992); 
R.E. Rudd, \NP {\bf B427} (1994) 81.}
\REF\polch{J. Polchinski, {\it Nucl. Phys.} {\bf B324} (1989) 123.} 
\REF\das{S.R. Das, S. Naik and S. Wadia, \Mod {\bf A4} (1989) 1033.}
\REF\david{F. David, {\it Mod} {\bf A3} (1988) 1651.}
\REF\distler{J. Distler and H. Kawai,  \NP {\bf B321} (1989) 509.}
\REF\god{P. Goddard, J. Goldstone, C. Rebbi and C. Thorn, \NP {\bf
B56} (1973) 109.}
\REF\tzani{R. Tzani, \PRD {\bf 38} (1988) 3112; \PRD {\bf 43} (1991) 1254.}
\REF\sakita{J. Gervais and B. Sakita, \PRL {\bf 30} (1973) 719.} 
\REF\mande{S. Mandelstam,  \NP {\bf B64} (1973) 205.}
\REF\deser{S. Deser and R. Jackiw, preprint MIT-CIP-2469, Oct. 95, 
hepth/9510145.}
\REF\sak{B. Sakita, in {\it Strings, Lattice gauge Theory, High Energy
Phenomenology}, proceedings of the Winter School, Panchgani, India,
1986, edited by V. Singh and S. Wadia (World Scientific, Singapore,
1986).}
\REF\mayer{R.C. Myers, \PL {\bf B199} (1987) 371.}
\REF\tzan{R. Tzani, \PRD {\bf 39} (1989) 3745.}
\REF\kato{M.Kato, preprint UT-Komaba/95-12, Dec. 1995, hep-th/9512201.}


\def\a{\alpha}

\def\d{{\delta}}

\def\r{\rho}
\def\t{\tau}
\def\e{\epsilon}

\def\s{\sigma}
\def\noi{\noindent}

\noi
{\bf I. Introduction}

The two-dimensional string theory with a linear background charge term
is known to have a non-trivial spectrum, that is, an infinite number
of physical states at discrete values of the momentum.
These states, originally discovered in the
$c=1$ matrix model [\gros], have been well analyzed in this theory in the
conformal gauge [\klebanov, \dani, \lian, \witten, \bouw]. 
In the light-cone gauge, however, one naively expects
that the two-dimensional theory should have no degrees of freedom left,
other than the tachyon, since there are no transverse degrees of
freedom in two dimensions and the light-cone gauge fixes completely the
longitudinal components of the theory. Then the question which arises
is how the light-cone gauge can see these extra states which are of
longitudinal nature.

This is related to the question of the validity of the light-cone
gauge itself. The main question we would like to address is the
following: Is the light-cone gauge a physical gauge, in the sense
that it gives the physical spectrum of the theory, or is it a 
non-legitimate choice which results in a truncation of the theory? 

A related issue is the difference in the spectra obtained from 
BRST and other quantization procedures. It is widely agreed that
BRST quantization gives the most trustworthy approach to the
construction of the quantum string spectrum. However, one expects
to obtain the same physical spectrum in the 
path-integral quantization procedure as well. All known studies 
on the subject have been done in the operator and BRST formalism of
the theory. We are interested to study the problem in the
path-integral formalism in an analogous fashion as Weinberg did for
the vertex operators in string theories [\wien]. Specifically, 
we would like to observe how
the discrete states are realized in Polyakov's path-integral formulation. 
                  
There have been two main works which address the question of the
discrete states in the light-cone gauge in this theory.
One is by Smith, who has fixed the light-cone gauge
conditions $ g _ {--} = 0 $ and $ X ^ + = p ^ + \t$ independent of the
conformal gauge [\smith]. He also fixes the $ g _ {+-} $ component of the
metric to one, using a modified Weyl symmetry (the pseudoWeyl
symmetry) of the theory. He studied the BRST cohomology of this 
light-cone gauge fixed theory and found no discrete states using 
this procedure. 

The other independent approach is that of Rudd, who fixes the
light-cone gauge condition after fixing the conformal gauge, 
using the residual symmetry left in the theory after imposing
the conformal gauge conditions [\rud]. He needs to fix the $ X ^ + $
component everywhere but at the special values of momentum in
order to avoid singularities in the gauge fixing condition. In this
way the corresponding modes of $ X ^ \pm $ remain in the 
physical Hilbert space. He obtains a Lagrangian formulation of
the gauge-fixed theory that explicitly depends on the oscillators 
(modes) of the $X ^ +$ component. He is using functional 
integration methods for his gauge fixing, but his analysis of the spectrum 
is based on the BRST charge operator.  

In this paper we will investigate the light-cone
gauge in Polyakov's path-integral formulation of the theory.
The analysis will be exactly at the same level as in the conformal
gauge. Specifically,
we will fix only the two-dimensional reparametrization invariance
using the light-cone gauge conditions, as in
the conformal gauge, where one sees the discrete
states before fixing the residual symmetry. The residual symmetry is
interelated with the physical spectrum of the theory and fixing it
from the start
can overconstrain the system. The other point we would
like to make is that we are not fixing the $ g_{+-}$ component of the
metric at this point, the main reason being that the (pseudo)Weyl
symmetry, used to fix this condition in the work of Smith, is not really
a symmetry at the classical level of the theory. Therefore one does not
know a priori how to treat a symmetry which is not classically there. 
Leaving the $ g _ {+-} $ component free we can study the conformal
invariance and anomaly of the theory in the path-integral formulation.

Before we start our analysis let's point out that the
conformal gauge approach which is analyzed in the literature 
uses the one-dimensional string theory as its starting point. 
It has been shown that the one-dimensional non-critical string theory
is equivalent to the two-dimensional critical string
with a background charge of value equal to $2 \sqrt 2$ [\polch, \das]. 
In that approach the $ g _ {+-} $
component of the metric becomes the Liouville field which plays the
role of an extra string dimension in the two-dimensional theory. It is
not obvious, however, how to fix the light-cone gauge starting from
the one-dimensional theory, since there are not even enough degrees of
freedom to make a light-cone component. Moreover, this equivalence
stems from the work of David, and Distler and Kawai [\david, \distler]
which is based on
a non-trivial anzatz. One does not know that similar arguments will
give results in the light-cone gauge as well.

Therefore in this paper we study the light-cone gauge starting
from the two-dimensional string with a background charge term, as in
ref. [\smith]. Our
main interest in this paper is to establish the equivalence between
the light-cone and the conformal gauge fixing of the theory in the
path-integral formulation of the theory. The
light-cone gauge we use is in the same spirit as the original light-cone of
Goddard, Goldstone, Rebbi and Thorn [\god], that is, it is fixed
independently of the conformal gauge [\tzani].
 
The plan of this paper is as follows. In section II, we fix the
light-cone gauge in the path-integral formulation of the theory. 
In section III, we quickly review the conformal gauge in the
same formalism, in order to compare the results with the light-cone
gauge ones. Section IV discusses the determinants in the light-cone
gauge. (For completeness we give the details of the calculation of
the determinant in an appendix.) It is shown explicitly that in
this gauge the theory is conformally invariant for the value of the
background charge $ Q = 2 \sqrt2 $. 
Also by comparing the gauge-fixed expressions of the light-cone 
and conformal gauge we show the equivalence of the two theories
at the generating functional level. Then we discuss the zero modes 
of the ratio of determinants in the light-cone gauge and their
relevance to the discrete states. 
Finally, section V contains the conclusions. 

{\bf II. Light-cone gauge fixing}

We start from the following action for the two-dimensional string with a
linear background charge term:
$$
S= -{1 \over {4 \pi \a ^ \prime}} \int d ^ 2 \sigma ( {\sqrt g} g ^{ab}
\partial _a X ^ \mu \partial _ b X _ \mu + \a  ^ \prime n \cdot X
\sqrt g R ^ {(2)})
\eqn\ac$$
where $\sigma = (\sigma ^0 , \sigma ^ 1)$ are coordinates in the
two-dimensional parameter space. 
$X ^ \mu (\sigma)$, $\mu = 1, 2 $ defines a map
from the two-dimensional parameter space to a surface in the
two-dimensional physical space-time. The metric $g _ {ab}$ characterizes
the parameter space. $ n ^ \mu = (n ^1 , 0) $ is a space-time vector, 
$\a ^\prime$ is the inverse string tension and $R ^ {(2)}$ is the Ricci 
scalar.

This action is invariant under two-dimensional reparametrizations while 
the Weyl symmetry is broken by the background charge term. 
One can recover the Weyl symmetry, however, by scaling the 
$X ^ \mu$ field accordingly, that is, by making  
the following transformations: 
$$
\eqalign{ \delta g _ { ab } &= \epsilon g _ {ab} \cr
\delta X ^ \mu &= - { \alpha ^ {\prime} n ^ \mu \epsilon \over 2 } \cr
\delta R ^ {(2)} &= -\epsilon R ^ {(2)} - { 1 \over {\sqrt g}} \partial _  
a ( \sqrt g g ^ {ab} \partial _ b \epsilon) ~. \cr
}
\eqn\we$$    
It is easy to see that this transformation leaves the classical action
invariant only to order $\a ^ \prime$. Terms of order ${\a  ^ \prime }^2 $
break this symmetry. Despite the fact that this is not a symmetry of
the classical action it has been used 
in the literature as such in order to fix the third component of the metric
[\smith, \rud]. In this paper we will take the point of
view that this is not a symmetry of our action and
therefore we will not fix it. In order to make our argument more clear, 
we will analyze both the conformal and light-cone gauge in parallel.
The gauge conditions will be
fixed in both cases using the reparametrization invariance of the
action. Our formulation is Polyakov's path-integral one.
We also start with a general background charge which will be specified
later.
 
In the rest of the paper, mainly for calculational convenience, but 
also because it makes our discussion more transparent, we will use the
following redefined field variables, as in [\smith]
$$
\eqalign{\tilde g ^ {ab} &= \sqrt g g ^{ab} \cr
\tilde X ^ \mu &= X ^\mu + { \a ^ \prime n ^ \mu \over 2 } \ln \sqrt g~. \cr }
\eqn\red$$
The new fields are invariant under the Weyl transformations \we. 
In terms of these new variables the action takes the following form
$$
S= -{ 1 \over 4 \pi \a ^ \prime } \int d ^2 \s [ \tilde g ^ {ab}
\partial _ a \tilde X _ \mu \partial _b \tilde X ^ \mu + \a ^ \prime 
n \cdot \tilde X \tilde R ^{(2)} - {{ \a ^ \prime } ^ 2  n ^ 2 \over 2 } 
\ln \sqrt
g ( \tilde R ^{(2)} - { 1 \over 2 } {\tilde \Delta } \ln \sqrt g ) ] 
\eqn\nac$$
where $ \tilde R ^ {(2)}= \sqrt g R ^ {(2)} + \sqrt g \Delta \ln \sqrt
g$ and $ 
{\tilde \Delta } = \partial _ a ( \tilde g ^ {ab} \partial _ b ) $.
Notice that the action \ac\ cannot be re-expressed completely in terms of
the Weyl invariant fields \red. The presence of the $\sqrt g $ in the terms of 
order $ {\a ^
\prime } ^ 2 $ in \nac\ reflects the fact that the action is not invariant
under Weyl transformations. 

Polyakov's path-integral formulation of the Virasoro
amplitude is given by 
$$ 
\int \cdot \cdot \cdot \int  \prod _ j d ^ 2 \s _ j {\sqrt g (\s _ j)
} {\cal D } g _ {ab} {\cal D } \tilde X ^ \mu 
e ^ { iS~ +~i \int d ^2 \s j ^ \mu \tilde X _ \mu} 
\eqn\po$$
where $S$ is given by \nac\ and $j ^ \mu$ is an external source term
defined by
$$
j ^ \mu (\s) = \sum _ j \delta ^ 2 ( \s - \s _ j ) k ^ \mu _ j ~.
\eqn\st$$
The reparametrization-invariant measure in \po\ is defined by 
$$
\eqalign {
{\cal D } g _ {ab} &= \prod _ {\sigma} d g _ {++} d g _ {+-} d g _ {
--} (g ) ^ { - 3/2} \cr
{\cal D } X ^ \mu &= \prod _ {\s} d X ^ + d X ^ - ~.\cr} 
\eqn\dm$$
(Notice that from now on we drop the tilde notation from the $X ^ \mu$
variable in order to simplify the notation.)
We use the Minkowski metric for the parameter space. In writing the 
integral representation of Virasoro amplitude \po, we assume that we
have divided by the M\''obius volume and the reparametrization volume 
(diffeomorphism), since the integrand is invariant under
two-dimensional general coordinate transformations.

We choose the light-cone gauge conditions as in ref. [\tzani], that is,
$$
\eqalign{g _ {--} &= 0 \cr
X ^ + &= f (\s) ~.\cr }
\eqn\lc$$
The first of these conditions corresponds to $ ( \dot X - X ^
\prime ) ^ 2 = 0 $ in the Nambu-Goto Lagrangian and
$ f (\s) $ is an arbitrary function of the two-dimensional $\s $. 
This is the gauge originally chosen by Gervais and Sakita
in their study of the interacting string [\sakita].

We fix the gauge using the invariance of the action under the general
coordinate transformations in two-dimensional parameter space
$$
\s ^ a \rightarrow \s  ^ {  (\e) a } = \s ^ a + \e ^ a (\s) ~.
\eqn\rt$$
Under infinitesimal reparametrization transformations the variables $
 X ^ \mu $ and $ g _ {ab} $ transform as
$$
\eqalign{  X ^ \mu &\rightarrow  X ^ {(\e) \mu } (\s) 
=  X ^ \mu (\s)- \e ^ a \partial
_ a  X ^ \mu (\s) - { \a ^ {\prime} n ^ \mu \over 2 } \partial _ a \e ^ a
\cr  
g _ {ab} &\rightarrow g _ {ab} ^ {(\e)}= g _ {ab} - \e ^ c \partial _
c g _ {ab}
- \partial _ a \e ^ c g _ {cb} - \partial _ b \e ^ c g _ {ac} ~. \cr }
\eqn\as$$
According to the Faddeev-Popov procedure, we change the integration variables 
$ d X^ +$ and $ d g _ {--}$ to the
two parameters of the transformations $ d \e ^ + $ and $ d \e ^ - $. 
The light-cone notation that we use throughout this paper is 
$$
\eqalign{\s ^ {\pm} &= { 1 \over \sqrt 2 } ( \s ^ 0 \pm \s ^ 1 ) \cr
g _ {\pm \pm } &= {1 \over 2 } ( g _ {11} + g _ {00} \pm 2 g _{ 01} )\cr
g _ {+-} &= { 1 \over 2} (g _ {11} - g _ {00} ) ~. \cr}
\eqn\no$$

Next we insert in the integral the identity
$$
\int D \e ~ \Delta _ {FP} (X, g) \prod _ \s \d (X ^ { +(\e)} - f (\s) ) 
\prod _ \s \d (g _ {--} ^ {(\e)} ) =1 
\eqn\id$$
where $ D \e$ denotes the invariant measure of the transformation and
$\Delta _ {FP}$ is the Faddeev-Popov determinant given by
$$
\Delta _ {FP} = {\rm det} \pmatrix{-\partial _ + f - { \a ^ \prime n ^
+ \over 2
} \partial _ + &- \partial _ - f - { \a ^ \prime n ^ - \over 2 }
\partial _ - \cr
-2 g _ {+-} \partial _ - &0  \cr}~ .
\eqn\fp$$

The next step is to make an inverse reparametrization transformation 
to all variables in
the path-integral. Since the integrand, measure and Faddeev-Popov
determinant are invariant under the transformation, the
integration over the parameter $\e$ can be factored out as a
constant equal to the volume of the reparametrization group. 

The integrations over $X ^ + $ and $ g _ {--} $ are trivially carried
out and the gauge fixed action is given by
$$
\eqalign{S = -{1 \over {4 \pi \a ^ \prime }} &\int d ^2 \s \left[
2\tilde g ^ {--} \partial _ - X ^ - \partial _ - f + 4
\partial _ + f \partial _ - X ^ - - \a ^ {\prime} n ^ + X ^ - \partial _
- ^ 2 {\tilde g} ^ {--} \right. \cr
&\left.  
- \a ^ \prime n ^ + f \partial _ - ^ 2 {\tilde g} ^ {--}
+ {\a ^ \prime} ^ 2 {n ^ +} ^2 \r [ \partial _ - ^ 2 {\tilde g} ^ {--}
+ \partial _ + \partial _ - \r +{1 \over 2 } \partial _ - ( {\tilde g}
^ {--} \partial _ - \r )] \right] \cr
& + i \int d ^ 2 \s
\left[   j ^ + X ^ - +  j
^ - f (\s) \right] \cr }
\eqn\gfl$$
where we have used the fact that $n ^ -= n ^ +$ and $\r$ stands for $
\ln \sqrt g$. 
Notice also that $ g ^ {+-} = - { 1 \over |g| } g _ {+-} $ and $\tilde g ^
{+-} = \sqrt g g ^ {+-} = 1 $ in this gauge.

Next we choose the function $f(\s)$ such that it satisfies the
following equation:
$$
\partial _ - \partial _+ f (\s) = - \pi \a ^ \prime j^+ (\s) ~.
\eqn\eq$$
With this choice of $f(\s)$ the linear terms in $ X ^- $ in \gfl\ cancel.
The solution of \eq\ is given by 
$$
f (\s) = \a ^\prime  \sum _ j k _ j ^+ \ln | \s - \s _ j | ~.
\eqn\lap$$
The integration over $ X ^-$ in the integral \po\ is now easily
carried out and results in
the following inverse determinant and delta function for $ \tilde g ^{--}$ 
$$
\left[ det ^ \prime \left( \partial _ - f \partial _ -  - { \a ^\prime n
^+ \over 2 }
\partial _ - ^2 \right) \right] ^{-1} \delta (\tilde g ^{--} )
\eqn\re$$
where by prime we mean that the determinant does not contain the
zero mode.
Because of the delta function we can now perform the integration over 
$g _{++}$. We obtain
$$
\int \cdot \cdot \cdot  \int \prod _ k  d ^ 2 \s _ k 
d \r 
{ \Delta _ {FP} \over {det ( \partial _ - f
\partial _ - - {Q \over 2 \sqrt 2 } \partial _ - ^ 2 ) } }
e ^ {{ i Q ^2 \over 8 \pi \a ^ \prime } \int d ^ 2 \s  
\partial _ + \r \partial _ - \r -  \sum _ i 
\sum _ {j \neq i } k ^ - _ i k ^ + _ j ln |\s _ i - \s _ j |
 }
\eqn\fl$$
where we have substituted $ e ^ \r $ for $ g _ {+-}$ and $Q$ for $ \sqrt 2 \a 
^ \prime n ^ + $. From now on we omit the prime from the
determinant. The exclusion, however, of the zero modes in the last
ratio of determinants is to be understood. 

The last expression is the light-cone gauge-fixed path-integral that
will concern us in the rest of this paper. In this expression, however,  
all variables and determinant operators are defined on
the two-dimensional world sheet. In order for this to coincide with
Mandelstam's formulation of strings, one has to perform Mandelstam's
mapping [\mande]. This procedure is well known and analyzed 
in a similar context in ref. [\tzani], but for completeness
we shall give at this point a brief discussion for this specific case. 

We first have to perform the following Wick rotation
$$
\s ^ 0 = i \xi ^ 2 ~~, ~~~~~~\s ^ 1 =\xi ^ 1 ~.
\eqn\wr$$
Under this rotation the path-integral becomes
$$
\int \cdot \cdot \cdot \int \prod _ k d ^2  z _ k d \r 
{ \Delta _ {FP} \over {det ( \partial_ {\bar z}f 
\partial _ {\bar z} - {Q \over 2 \sqrt 2 } \partial _ {\bar z}  ^ 2 ) } }
e ^ {{ Q^2 \over 8 \pi \a ^ \prime } \int d ^ 2 \xi  
\partial _ z \r \partial _ {\bar z} \r - \sum _ {i
\neq j } k ^ - _ i  Re {\cal F} (z _ i) }
\eqn\aw$$
where
$$
{\cal F } (z) = \sum _ j k _ j ^+ ln ( z - z _ j ) 
\eqn\cf$$
and $z$ stands for $ z$ and $\bar z$ since $ z= \xi ^ 1 + i \xi ^ 2$
and $\bar z = \xi ^ 1 - i \xi ^ 2 $. In this expression the variables
$\xi$ and $z$ are defined on the entire complex plane. 

Then we make Mandelstam's conformal transformation
$$
z \rightarrow w = {\cal F }(z) \equiv \tau + i \s ~. 
\eqn\ae$$
This defines a map from the entire plane to Mandelstam's tube. 
Under this transformation our gauge fixed function $f (\s)$ becomes
the usual light-cone gauge condition $\tau$. The
points $ z _ i$ transform into strings $i$
at initial and final times $ \tau = {\mp} \infty $, depending on the
values of $k _ i ^ +$ , $k _ i ^ + > 0 $ or $ k _ i ^ + < 0 $ for
incoming or outgoing strings, respectively. The last exponential
factor of \fl\ becomes
$$
exp ( - \sum _ i k _ i ^ - \tau _ i ) ~.
\eqn\lf$$
where $ \t _ i = \sum _ { j \neq i} k ^ + _ j ln | z _ i - z _ j |
$.
Then we perform the
same transformation on the $\r$ integration variable and the
determinants such that all variables are defined on Mandelstam's tube.
The result for the path-integral is now
$$
\int \cdot \cdot \cdot \int \prod _ k d {\tilde \t }_ k d \r 
{ \Delta _ {FP} \over {det ( \partial_ {\bar w}f
\partial _ {\bar w}  - {Q \over 2 \sqrt 2} \partial _ {\bar w} 
 ^ 2 ) } }
e ^ {{ Q ^ 2 \over 8 \pi \a ^ \prime } \int d ^ 2 w  
 \partial _ w \r \partial _ {\bar w} \r -  \sum _ i k ^ - _ i 
\t _ i }
\eqn\az$$
where $ \tilde \t _ k$ are the interaction times of the string.
Note that the exponent \lf\ does not depend on the variable $w$ and 
therefore comes out of the $w$-integral. 
However, this exponential factor implicitly depends on the interaction
times $ \tilde \t _ k $ since changing $ \t _ i $ reflects a change in
both $ \t _ i$ and $ \tilde \t _ i$, which are defined by
$$
\tilde \t _ i = {\cal F } ( \tilde z _ i ), ~~~{\rm where}~~~ {
\partial {\cal
F} (z) \over \partial z } |_ {z = {\tilde z _ i }} = 0 .
\eqn\sk$$ 
Thus the integration over the $ \tilde \t _ k$ variables makes this 
factor non-trivial and therefore the dependence of the path-integral
on both $ k ^ - $ and $ k ^ + $ essential. 

The expression \az\ gives 
Mandelstam's picture for two-dimensional strings with a background
charge. Since there are no transverse degrees of freedom in this
theory, the exponent is essentially the factor \lf. 
The determinants depend on the $ g _ {+-}$ component
of the metric, which is a result of the non-invariance of the measure
under conformal symmetry. The $g _ {+-}$ (or otherwise $\r$)-dependence 
of the integrand defines the conformal anomaly. In this theory, due to
the fact that the background charge is a quantum correction of the
classical action, there is also a $\r$-dependence in the exponent of
the integrand and hence an extra contribution to the conformal anomaly
due to this term. A straightforward way to see this is by using the
action \nac. The change of the action under Weyl transformation 
gives essentially the result 
$$
\int  d ^ 2 \s \sqrt g R ^ {(2)} \e 
\eqn\wv$$
which is nothing but the Weyl anomaly term. 
In this sense the background term charge plays the
role of an effective (WZW-like) action for the conformal anomaly
[\deser]. This fact will be explicit from our calculations below.

The anomaly in the light-cone gauge will be given 
in section IV. In the next
section, in order to compare our results with the results of the 
conformal gauge, we briefly analyze the conformal gauge in this theory.

{\bf III. Conformal gauge fixing.}

We start again from Polyakov's path-integral expression, given by \po,
and we fix the two-dimensional reparametrization invariance of the
action, as before.

The conformal gauge conditions are now
$$
g _ {++} = 0 ~~{\rm and} ~~g _ {--} = 0  
$$
where, again, we use light-cone notation.
 
We use the Faddeev-Popov method for the gauge fixing. The Faddeev-Popov
determinant is now an essentially diagonal matrix given by 
$$
\Delta _ {FP} = \pmatrix{ &0 & -2 g _ {+-} \partial _ + \cr
                          & -2g _ {+-} \partial _ - &0 \cr } ~.
\eqn\fpc$$
After trivially integrating over $ g _ {++} $ and $g _ {--}$ in the
path-integral, we obtain 
$$
\eqalign{
\int \cdot \cdot \cdot \int \prod _ k d ^ 2 \s _ k {\sqrt g (\s _ k) }
d \r d X ^ + d X ^ - \Delta _ {FP} &e ^ {
- {i \over \pi \a ^\prime } \int d ^2 \s \left( \partial _ +
X ^+ \partial _ - X ^- - {Q ^2 \over 8 } \partial _ + \r \partial 
_ - \r  \right) } \cr 
& e ^ { i \int d ^ 2 \s \left( j ^ + X ^ - + j ^ - X ^ + \right)} ~. \cr}
\eqn\cgf$$
This is the path-integral expression for the conformal gauge fixed
theory. The source term is defined again by \st.
In order to obtain the determinants 
and extract the $ \r $-dependence of the part of the action 
containing the matter fields, we perform the integrations over $ X ^ +
$ and $ X ^ - $. Because of the presence of the source term, 
the integrations are done as follows: 
we first integrate over the $ X ^- $ field. The result is the following
delta function
$$
\delta ( \partial _ + \partial _ - X ^+  + \pi \a ^\prime j ^+ ) ~.
\eqn\eqc$$
This, then, means that the integration over $ X ^+ $ replaces the $ X ^+ $
variable in the action with the solution of the following equation 
$$
\partial _ + \partial _ - X ^+ = - \pi \a ^ \prime j ^+ (\s)
\eqn\so$$
and gives the following inverse determinant 
$$
\left[ det ( \partial _ + \partial _ - ) \right] ^{-1} .
\eqn\dt$$ 
Comparing at this point \so\ with the equation
\eq\ of the previous section, we see that the variable $X ^ +$ of this
gauge satisfies the same equation as the function $f (\s)$ of the
light-cone gauge conditions. The meaning of this is that 
after the integration over $X
^ -$, the $ X ^ + $ field of the conformal gauge-fixed theory takes the
value of the light-cone gauge fixed theory. 

Then the expression \cgf\ becomes
$$
\int \cdot \cdot \cdot \int \prod _ k d ^ 2 \s _ k
d \r  { \Delta _ {FP} \over 
det ( \partial _+ \partial _ - ) }
e ^{{ iQ^2 \over 8\pi \a ^\prime } \int d ^2 \s  
(\partial _ + \r \partial _ - \r)~ -~  \sum _ {i \neq j} k
^ - _ i f ( \s _ i)  }
\eqn\lcg$$
where in the last expression we have called $f (\s)$ the solution of the
equation \so\ in an obvious notation.

Comparing the last expression with \fl\ we see that they are the
same except for the determinant-ratio factor. 
The calculation of the determinants of the operators in the conformal gauge is 
known. Therefore, we only state the results. They are given by
$$
det (\partial _ + \partial _ - ) = e ^ {- {i \over 12 \pi \a ^ \prime
} \int d ^2 \s \partial _ + \r \partial _ - \r }
\eqn\ddc$$
$$
\Delta _ {FP} = e ^{- { 26 i \over 24 \pi \a ^ \prime} \int d ^2 \s 
\partial _ + \r \partial _ - \r } ~. 
\eqn\dnc$$
Putting now \lcg\ together with \ddc\ and \dnc\ we see that the anomaly
in the conformal gauge cancels for
the value of the background charge $ Q = 2 \sqrt 2 $.

In the next section we shall discuss these determinants in the light-cone
gauge. We shall show that the contribution to the anomaly coming from
the determinants and the contribution coming from the background
charge term again cancel for the same 
value of the background charge, that is, for $ Q = 2 \sqrt 2$. 

{\bf IV. The Determinant in the light-cone gauge}

The Faddeev-Popov determinant in the light-cone gauge and the determinant 
coming from integrating out the matter field variables are 
combined in the following ratio
$$
{{ det \pmatrix{  -x _ + - { Q \over 2 \sqrt 2 } \partial _ -  & 
- x_- - {Q \over 2 \sqrt 2 }  \partial _ - \cr
               -2g _ {+-} \partial _ - & 0 \cr } } \over 
{det \left( x _ - \partial _ - - { Q \over 2 \sqrt 2 }
\partial _ - ^ 2 \right) } }
\eqn\ad$$
where we have denoted $ \partial _ - f $ by 
$ x _ -$. 

In order to calculate this determinant we need to regularize the 
functional integrations associated with it and extract the 
$\r$ dependence from the ratio \ad\ .
We use the Pauli-Villars regularization procedure [19]. 
Since this method is well analyzed in ref. [14], we here state
only the result. (For completeness we include the calculations 
of the determinant in the appendix A.)
It is given by 
$$
{  \Delta _ {FP} \over \det (x _ - \partial _ - - { Q \over 2
\sqrt 2 } \partial _ - ^ 2 ) }= exp  \left[-  {24 \over 24 \pi 
\a ^ \prime}  \int d ^ 2 \xi [{ 1 \over 2 } ( \partial _ + \r
\partial _ - \r ) + \mu ^ 2 e ^ \r ] \right]
\eqn\xx$$

Due to the form of the determinant in the light-cone
gauge, one, naively, would expect that the result 
depend explicitly on $Q$. However, because of cancelations between the
numerator and denominator the result does not depend on the background charge
$Q$. 
 
It is now easy to see that the total $\r$-dependence on the integrand in the
expression \az\ cancels for the value of the background charge $ Q
= 2 \sqrt 2 $. 

The conclusions from the last results are: first, that the 
two-dimensional string theory with a background charge is conformally (scale)
invariant at the quantum-mechanical level for the specific value of the
background charge $ Q= 2 \sqrt 2$. Even if this result has been
known in the conformal gauge for some time 
(see for instance ref. 
[\polch, \distler]), in this paper we have made it more clear in both 
gauges\footnote\dag{A light-cone gauge canonical
approach has been also analyzed by Myers in the theory of
$d$-dimensional string with background charge term [\mayer].}.
We have explicitly shown that the background charge
term in the original action is nothing but the Wess-Zumino term for
the anomaly, since the contribution coming from that term cancels
exactly the one coming from the ratio of determinants for the
specific value of Q. The second conclusion obtained from the above 
analysis is that the light-cone gauge-fixed 2-d string theory is 
equivalent to the
conformal gauge-fixed one. The path-integral expressions \fl\ and
\lcg\ coincide except for the ratio of determinants. In fact, since 
we have integrated out all the fields except $\r$, in our approach, 
this determinant factor is essentially the content of the
theory. In this section we showed that the determinants
coincide as well, showing the complete equivalence of the two
gauge-fixed theories. 

Then, since the two theories have been shown to be equivalent at the 
partition function level, one would conclude that the
spectra of the theories are the same. Therefore the light-cone gauge
must contain the physical states that are present in the conformal 
gauge. In the path-integral approach, however, naively it seems that,
after integrating out the $ X ^ \pm$ variables and calculating 
the determinant, 
nothing which has some physical content is left in the theory. See, 
for instance, the expression \az\ in the light-cone gauge. 
Therefore any physical states contained in 
the theory must originate from the factor $ {\rm exp} { (\sum _ i k ^ - _ i 
\tau _ i)} $ and the zero modes of the determinant. (Remember 
that the zero modes of the determinant were excluded in our previous
discussion.) We shall briefly discuss the zero modes of the
determinants next.

The zero modes of the Faddeev-Popov determinant correspond to the
residual gauge transformations left in the theory after fixing the
reparametrization invariance. These are given by
$$
\eqalign{ 
\e ^ + &= \e ^ + ( \s ^ + ) \cr
\e ^ - &= - ( \e ^ + + 2  {\e ^ \prime } ^ +) + \tilde g ( \s ^ +) e ^
{ - {1 \over 2} \s ^ - } \cr
}
\eqn\zmn$$
where by prime we mean derivative with respect to the argument and the
$\tilde g $ here is an arbitrary function of $\s^ +$ only.
 
On the other hand the zero modes of the determinant in the denominator
correspond to the unphysical degrees of freedom and 
are given by
$$
\Psi = 2 \pi (\s ^ + ) + \phi ( \s ^ +) e ^ { - {1 \over 2} \s ^ - }
\eqn\zmd$$
where $\pi$ and $\phi$ are arbitrary functions of the argument.
These ``residual'' degrees of freedom are the components of
the gauge field which change under the residual transformations and 
are eliminated 
upon fixing the residual gauge transformation. If the theory contained
no discrete states, they should be exactly
as many as the number of the residual gauge transformations. 
(In practice the elimination of the unphysical degrees of freedom is
taking place by imposing on the states the physicality condition,
which is the generator of the residual symmetry.)
In this case, however, since extra states exist, there exist some
residual gauge transformations which do not change the
gauge fields. These gauge transformations do not eliminate unphysical
degrees of freedom. Therefore for the corresponding values of momenta 
extra states remain in the theory. (In the case of QED, for instance,
the only degree of freedom  which remains this way in the theory is
the trivial constant mode, which exists at zero momentum.)
In terms of the zero modes of the determinants, then, the expectation 
is that
the Faddeev-Popov determinant should have some zero modes which do 
not correspond to zero modes of the determinant in the denominator. 
These extra zero modes are at the momenta at which the discrete
states exist.  

Comparing, however, the solutions of \zmd\ to the ones of \zmn\ 
we find that the only residual gauge transformation which does
not eliminate unphysical degree of freedom in this case corresponds 
again to the constant mode. 
A similar calculation in the conformal gauge gives exactly the same result. 
Only the constant mode survives from the determinants there as well. 
This result is expected, since in our path-integral formalism we have 
included only the tachyon vertex operator. The discrete states 
corresponding to higher spin particles should arise in a similar way
by including the appropriate vertex operators in the path-integral. 
Our present result, therefore, provides 
an indication that the discrete states may lie in the
zero modes of the determinants, and this is not particular to the
light-cone gauge. This is one main observation in this paper.

\bigskip\noi
{\bf V. Conclusions }

In this paper we have studied the light-cone gauge in 
two-dimensional string theory with a linear background charge term in the
path-integral formulation. One of the aims of this work is to
understand the presence of the background charge
in the Lagrangian and the role of it in the conformal invariance of the
theory in light-cone gauge. This question becomes relevant
since classically the background charge term breaks Weyl symmetry. One
therefore would like to understand how the conformal symmetry is
restored at the quantum mechanical level. We have shown that one can
fix the reparametrization invariance by choosing the light-cone gauge 
conditions in this theory independently of the conformal gauge.
The gauge-fixed theory then depends on the conformal component of the 
metric and in general it is not conformally invariant. By requiring 
conformal invariance of the theory at the quantum-mechanical level, 
we have shown that the conformal anomaly cancels when the value of the
background charge equals $ 2 \sqrt 2$. In particular, the dependence
on the conformal factor coming from the ratio of determinants cancels the
dependence coming from the background charge term exactly at that
value of the background charge. This is an explicit realization of the
fact that the background charge term plays the role of the Wess-Zumino
term in the Lagrangian [\deser]. 

Moreover we have shown that the light-cone gauge-fixed theory obtained
by imposing the light-cone conditions from the beginning, that is,
independently of the conformal gauge, is equivalent to the theory
one obtains in conformal gauge\footnote\dag{Similar results, which
show the equivalence between the conformal gauge and Polyakov's
light-cone gauge in this theory, have been recently obtained by Kato
using a similarity transformation in the BRST charge [\kato].}. 
This has been shown
by comparing the two path-integral expressions after fixing
the conformal and light-cone gauges respectively. The calculation of the
determinants in both gauges give the same result. 

Finally, let's emphasize that the ultimate goal of this approach is to
understand the existence of the discrete states in the light-cone
gauge in the path-integral formulation. This question has not yet 
been answered completely in this paper. 
Since, however, the conformal gauge-fixed and the
light-cone gauge-fixed theories are equivalent at the partition function 
level one expects that the spectra of the two theories are the same. 
Specifically, in our path-integral formulation the whole theory is 
contained in the ratio of the determinants and an exponential factor
which depends explicitly on the external momenta. From our analysis of
the zero modes of the determinants we have concluded that 
only the constant mode is hidden in the zero modes of the determinant 
in this formalism, which may correspond to the tachyon vertex operator 
present in our theory. This result  
is not a peculiarity of the light-cone gauge. The conformal gauge 
path-integral gives exactly the same results.

\bigskip\noi
{\bf Acknowledgments }

We would like to thank K.S. Stelle for
many discussions, encouragement and a critical reading of the manuscript, 
A.P. Polychronakos for useful discussions and a critical reading of
the manuscript, J. Schnittger for a helpful discussion and
K. Thielemans for conversations.
We would also like to thank R.E. Rudd for sending us his thesis and
for interesting remarks.

{\bf Appendix A. Calculation of the determinants}

In this appendix we discuss the calculation of the ratio of
determinants in the light-cone gauge.
Due to the fact that our gauge fixing is not diagonal, the determinant
operators are non-diagonal in this gauge. The components of the 
Faddeev-Popov determinant do not have the same signature. They both
operate on vector fields but give back scalars and two-tensors
respectively. In order to reduce the determinants to those of 
Laplacian (self-adjoint) operators we use a conformal invariant\footnote\dag
{By ``conformal invariance'' here we mean the symmetry which
transforms analytic functions into analytic and anti-analytic into 
anti-analytic.} regularization.

The best way to understand
the $\r$-dependence in the determinants is by using
Pauli-Villars regularization [\sak]. According to this procedure, the
regularization of the operator $ \nabla ^ 2 _ \xi $ is done as
follows: A number of auxiliary Bose or Fermi fields, with masses $M _
i$, are introduced in order to cut off the large momentum
contribution. The regularized determinant is then the determinant due
to all these fields:
$$
\prod _ i [ \det ( \nabla ^ 2 _\xi - M _ i ^ 2 e ^ { \r (\xi)} )] ^ {
C _ i}
\eqn\rd$$
where $ C _ i = \pm 1$ depending on the statistics, with $ C _
0 = 1$ and $ M _ 0 = 0$. The multiplication of the mass term with
$ e ^ {\r (\xi)} $ is done in order to have conformally invariant
operators. At the end of the calculation we set 
$$
1+ \sum _ i C _ i = 0 , ~~ \sum _ i C _ i M ^ 2 _ i = 0 ,~~{\sum _ i C
_ i M ^ 2 _ i \ln M _ i}_ {M _ i \rightarrow \infty} = {\rm finite}, 
\eqn\ddef$$
to assure that the result will be finite (regularized).
The $ \r$-dependence of the determinant is then 
computed  by 
$$
{ \delta \over \delta \r (\xi) } \ln \det {\nabla ^ 2 _ \xi } ^ {reg} 
= - \sum _ i C _ i M ^ 2 _ i e ^ { \r (\xi) } < \xi | { 1 \over
\nabla ^ 2 _ \xi - M ^ 2 _ i e ^ {\r ( \xi ) }} | \xi > ~.
\eqn\cd$$

In order to correctly calculate the quotient determinant \ad, one has to
regularize the ratio of determinants such that the infinities
between the numerator and denominator cancel. Also, the Faddeev-Popov
determinant is an infinite by infinite matrix and multiplication of
the determinants associated with these operators is not allowed in 
principle. It is, however, an
intractable problem to regularize the Faddeev-Popov
determinant as it is. In the calculation which follows we 
regularize the two operators of the Faddeev-Popov determinant
separately. In order to justify this we use the 
action formulation. 

Let's express 
the Faddeev-Popov determinant in terms of
ghost and antighost fields. It is given by 
$$
\Delta _ {FP} = \int d b d b ^{--} d c ^+ d c ^- e ^{ -i \int d ^ 2\s 
\sqrt g \left[ 
b ( x _ -  + { Q \over 2 \sqrt 2 } \partial _ - ) c ^- +
2 b ^{--} g _ {+-} \partial _ - c ^+ \right] } ~. 
\eqn\dg$$
Now we observe that the two terms in the action of \dg\ are expressed
in terms of two independent sets of fields, i.e. $ b ^ {--} , ~c ^ + $
and $ b, ~ c^ - $. Therefore they can be regularized in the action
independently, that is, by introducing two different Pauli-Villars
terms. This is equivalent to regularizing the determinants of the two 
operators in the numerator of \ad\ separately. In what follows we
shall regularize
the $\det (-2 g _ {+-} \partial _ - ) $ and $ \det (-x _ - - { Q
\over 2 \sqrt 2 } \partial _ - )$ independently. 

On the other hand, the bosonic part of the determinant can be
viewed as two independent determinants, i.e. $ \det (\partial _ - )$
and $ \det ( x _ - - {Q \over 2 \sqrt 2 } \partial _ - )$. Again this
stems from the fact that the relevant part of the action can be
written in first-order formalism by introducing auxiliary fields. 
It is 
$$
\tilde g ^ {--} (x _ - \partial _ - - { Q \over 2 \sqrt 2 } \partial ^
2 _ - ) X ^ - = \phi ^ {--} \partial _ - \xi _ -+ \psi (x _ - + { Q \over 2
\sqrt 2 } \partial _ - ) \lambda ^ - 
\eqn\ne$$
where $ ~\phi ^ {--},~ \xi _ - ,~\lambda ^ - $ and $ \psi$ are
auxiliary fields. Notice here that the action in first-order formalism
has been written in such a way that the operators involved act on
spaces with tensorial signatures similar to the ones of the ghost
action. 
One also finds that $ \det (x _ - + {Q \over 2 \sqrt 2 } \partial _ - ) =
\det ( x _ - - { Q \over 2 \sqrt 2 } \partial _ - )$, since if $ \psi
(x) $ is an eigenfunction of the first operator with eigenvalue
$\lambda$, then $ \psi (-x) $ is an eigenfunction of the second
operator with the same eigenvalue $ \lambda $ and the same boundary
conditions. Therefore, both operators have the same eigenvalues and
hence the same determinant. In what follows we 
regularize the bosonic and ghost-term operators using the method
described above.

We first compute the adjoints of the operators of the 
determinants. The determinants of the interesting operators are then 
computed by
extracting the square root of $\det O ^ \dagger O $, where $O$
stands for the relevant operator. In order to compute the adjoints of
the operators we must define appropriate inner products.
We define our inner products such that they are conformally covariant.
(We use
the $ g _ {+-} $ component of the metric in order to lower and raise
the indices.) 
The invariant inner product of two-tensor fields is defined
by
$$
(\Psi _ {++}, X _ {++} ) = \int d \s ^ + d \s ^ - g ^ {+-} \Psi ^ * _
{++} X _ {++} ,
\eqn\ip$$
where $ \Psi ^ * _ {++} = \Psi _ {--} $, while the inner products 
of two-scalars and two-vectors are defined by
$$
(\Psi, X ) = \int d \s ^ + d \s ^ - g _ {+-} \Psi ^ * X 
\eqn\sip$$
and
$$
( \Psi ^ + , X ^ + ) = \int d \s ^ + d \s ^ - g _ {+-} ^ 2 \Psi ^ {+*} X ^
+
\eqn\vip$$
respectively.

For the operators of the Faddeev-Popov determinant we have:
If $ P _ 1 = - x _ - - {Q \over 2 \sqrt 2 } \partial _ - $ the
adjoint $ P _ 1 ^ \dagger $ is defined by 
$$
(\delta X , P _ 1 \delta \e ^ - ) = ( P _ 1 ^ \dagger \delta X, \delta
\e ^ - ) .
\eqn\adj$$
Using the last relation and the definitions for the inner products we
obtain 
$$
P _ 1 ^ \dagger =  {g ^ {+-}} ^2 (-x _ + + { Q \over 2 \sqrt 2 }
\partial _ + ) g _ {+-}.
\eqn\ap$$

For $ P _ 2 = -2 g _ {+-} \partial _ - $ the adjoint $ P _ 2
^ \dagger $ is defined by
$$
(\delta g _ {--}, P _ 2 \delta \e ^ + ) = ( P _ 2 ^ \dagger \delta g _ {--} ,
\delta \e ^ + ) .
\eqn\ak$$
It is given by
$$
P _ 2 ^ \dagger = 2 {g ^ {+-}} ^ 2 \partial _ + .
\eqn\as$$

Similarly, we compute the adjoints for the operators in the
denominator of \ad. 
For $ T _ 1 = x _ - + {Q \over 2 \sqrt 2
} \partial _ - $ the adjoint is $ T _ 1 ^ \dagger = -{g ^ {+-}} ^ 2 
 (- x _ + + {Q
\over 2 \sqrt 2 } \partial _ + ) g _ {+-} $. 
For $ T _ 2 = \partial _ - $, the adjoint is $T ^ \dagger _ 2 = -
\partial _ + g ^ {+-} $. 

The dependence of these operators on $\r$ is computed by taking the
variation of  $\ln \det P _ 1 ^ \dagger P _ 1 $, $\ln
\det P _ 2 ^ \dagger P _ 2 $, 
$ \ln \det T ^ \dagger _ 1 T _ 1 $ and $ \ln \det 
T ^ \dagger _ 2 T _ 2 $ where 
$$
P ^ \dagger _ 1 P _ 1 =e ^ {-\r} (x _ + - { Q \over 2 \sqrt 2} \partial _
+  - { Q \over 2 \sqrt 2 } \partial _ + \r ) ( x _ - + {Q \over 2 \sqrt
2 } \partial _ - ) 
\eqn\eo$$
and 
$$
P ^ \dagger _ 2 P _ 2 = - 4 e ^ { -2 \r } \partial _ + (e ^ \r \partial
_ -)
\eqn\ew$$
while for the $ T$ operators it is 
$$
T ^ \dagger _ 1 T _ 1 = e ^ {-\r} (x _ + - { Q \over 2 \sqrt 2 }
\partial _ + - { Q \over 2 \sqrt 2 } \partial _ + \r ) 
(x _ - + {Q \over 2 \sqrt 2} \partial _ - )
\eqn\te$$
and
$$
T ^ \dagger _ 1 T _ 1 = - \partial _ + (e ^ { -\r} \partial _ -) 
\eqn\nt$$
where we have substituted $ e ^ \r $ for $g _ {+-} $.

Then the variation of the whole ratio of the determinants is given by
$$
\eqalign{
&\delta \ln { {\sqrt {\rm det} ( P _ 1 ^ \dagger  P _ 1)} {\sqrt {\rm det} 
( P _ 2 ^
\dagger P _ 2) } \over {\sqrt {\rm det} ( T ^ \dagger _ 1 T _ 1 ) } { \sqrt
{\rm det} ( T ^ \dagger _ 2 T _ 2 ) } }= \cr
& {1 \over 2} \left(\delta \tr \ln ( P ^ \dagger _ 1 P _ 1 ) + \delta 
\tr \ln ( P _ 2 ^ \dagger P _ 2) - \delta \tr \ln ( T _ 1 ^ \dagger T
_ 1 )  - \delta  \tr \ln ( T ^ \dagger _ 2 T _ 2 ) \right) \cr }
\eqn\to$$
Now, since the operators $ P ^ \dagger _ 1 P _ 1 $ and $ T ^ \dagger _ 1
T _ 1 $ are the same, they cancel out of the last expression and the 
variation of the ratio of the determinants is determined only
by the variation of $ P ^ \dagger _ 2 P _ 2 $ and $ T ^ \dagger _ 2
T _ 2 $. 

Following the regularization procedure given by \rd\ and \ddef, we
write the regularized quantities to be computed as 
$$
F _ 1 = \sum _ i C _ i tr \ln ( A _ 1 + M ^ 2 _ i e ^ \r)
\eqn\rga$$
and
$$
S _ 1 = \sum _ i C _ i tr \ln [ B _ 1 + M _ i ^ 2 e ^ \r ]
\eqn\se$$
where $A _ 1 = \partial _ + \partial _ - + \partial _ + \r \partial _ -$
and $ B _ 1 = (\partial _ + \partial _ - - \partial _ + \r \partial
_ - ) $ . 

In writing \rga\ and \se\ we have extracted the constant factor $\ln e ^
{ -\r} $.

The computation gives
$$
{ \delta F _ 1 \over \delta \r (\xi) } = - { 26 \over 24 \pi } \partial
_ + \partial _ - \r 
\eqn\cda$$
and 
$$
{\delta S _ 1 \over \delta \r (\xi) } = - { 2 \over 24 \pi } \partial _ +
\partial _ - \r ~. 
\eqn\sa$$

Putting now together the formulas \cda\ and \sa\ we obtain
$$
{  \Delta _ {FP} \over \det (x _ - \partial _ - - { Q \over 2
\sqrt 2 } \partial _ - ^ 2 ) }= exp  \left[-  {24 \over 24 \pi }  
 \int d ^ 2 \xi [{ 1 \over 2 } ( \partial _ + \r
\partial _ - \r ) + \mu ^ 2 e ^ \r ] \right]
\eqn\xx$$
where in the last expression we explicitly include the cosmological
constant term.

\refout
\vfill\eject

\end